# Evidence for Conical Magnetic Structure in M-type BaFe$_{12}$O$_{19}$ Hexaferrite: A Combined Single-Crystal XMCD and Neutron Diffraction Study

*Keshav Kumar, Shrawan K. Mishra, Sanjay Singh, Ivan Baev, Michael Martins, Fabio Orlandi, Pascal Manuel and Dhananjai Pandey\**

Mr. Keshav Kumar, Dr. Shrawan Kumar Mishra, Dr. Sanjay Singh, Prof. Dhananjai Pandey
School of Materials Science and Technology, Indian Institute of Technology (Banaras Hindu University), Varanasi, India-221005.
E-mail: dp.mst1979@gmail.com

Dr. Michael Martins, Dr. Ivan Baev
Universität Hamburg, Institut fü Experimentalphysik Luruper Chaussee 149, D-22761 Hamburg, Germany.

Dr. Pascal Manuel, Dr. Fabio Orlandi
Science and Technology Facilities Council Rutherford Appleton Laboratory ISIS Facility Didcot OX11 0QX, UK.

**Abstract**

The magnetic ground state of BaFe$_{12}$O$_{19}$ (BFO) was investigated using X-ray absorption spectra (XAS), X-ray magnetic circular dichroism (XMCD) and neutron diffraction studies at 1.2 K and 1.5 K, respectively. The XMCD measurements on single-crystals of BFO in grazing incidence geometry reveal canting of the spins away from the c-axis of the hexagonal unit cell. Single-crystal neutron diffraction studies reveal magnetic satellite peaks along the *00l* reciprocal lattice row around the forbidden $l = 2n \pm 1$ positions confirming conical type magnetic structure in the ground state of BFO. The observation of conical magnetic structure of BFO opens the possibility of type-II multiferroicity in undoped BFO also.

**Main Text**

Hexaferrites are technologically important and commercially used materials for a variety of applications such as permanent magnets, gyromagnetic devices, high-speed digital electronics, telecommunication, radar and energy generation.[1–7] Among the M, Y, W, Z, X and U-type of hexaferrites, recent years have witnessed a spurt of interest in Y[8–12] and Z-type[13–15] hexaferrites due to the discovery of type-II multiferroicity near the room temperature on account of conical magnetic structure in which the magnetic moment precesses about a specific crystallographic direction.[8,16] Y and Z-type hexaferrites exhibit longitudinal (//-c axis) conical ordering, confirmed by the observation of characteristics magnetic satellite peaks along the *hhl* reciprocal lattice rows around $l = 2n \pm 1$ positions which are otherwise forbidden by the nuclear and collinear magnetic structure.[17,18] On application of magnetic field transverse to the c-axis, these compounds exhibit transverse conical ordering which is at the heart of the type-II magneto-electric coupling through the inverse Dzyaloshinskii-Moriya (DM) effect.[18–20] The M-type hexaferrites, like barium hexaferrite $BaFe_{12}O_{19}$, have all along been believed to possess collinear magnetic order with moments aligned along the c-axis as per the Gorter model[21,22] as shown schematically in **Figure 1(a)**. In a recent work also, the collinear model was used to refine the magnetic structure of $BaFe_{12}O_{19}$ using single-crystal neutron diffraction data at 4K.[22] Since M-type hexaferrites occupy the major market share, there is considerable interest in discovering conical magnetic ordering and hence type-II multiferroicity in this family of hexaferrites also. In this context, it has been known that Sc doping in $BaFe_{12}O_{19}$ and $SrFe_{12}O_{19}$ induces longitudinal conical magnetic ordering whose cone angle about the c-axis increases with increasing Sc concentration.[23,24] More interestingly, these doped compositions have recently been shown to exhibit type-II multiferroicity with record-high ferroelectric polarization of ~ 25 $\mu C/m^2$ under transverse magnetic field.[25]

Since the existence of type-II magneto-electric coupling in hexaferrites is intimately linked with the existence of the conical magnetic order, it is of immense interest to search for conical ordering in undoped M-type hexaferrite also. The purpose of this letter is to present the first experimental evidence for the existence of such a conical ground state using, XMCD and neutron diffraction studies on single-crystals of $BaFe_{12}O_{19}$ at T ≲ 1.5 K. XAS and XMCD studies at the Fe $L_{2,3}$-edges confirm the non-collinear magnetic structure of $BaFe_{12}O_{19}$ at ~ 1.2 K. Single-crystal neutron diffraction studies at ~ 1.5 K, confirm conical magnetic order as revealed by the presence of magnetic satellite peaks along the *00l* reciprocal lattice row around the forbidden $l = 2n \pm 1$ positions.

The as-grown crystals of BaFe$_{12}$O$_{19}$ (BFO) exhibit hexagonal platelet shape as can be seen from inset (a) of **Figure 2** which depicts the image of one such single crystal. The lateral dimension of the crystal used is about ~ 4 mm while the thickness is ~ 0.8 mm. X-ray diffraction (XRD) pattern from powder obtained after crushing a few single-crystals shown in **Figure 2** confirms the crystalline nature of the crystals. There is no evidence for the common α-Fe$_2$O$_3$ impurity phase in the XRD pattern whose most intense peak occurs at $2\theta = 33.2^0$.[26] All the peaks in the diffraction pattern were indexed using Rietveld refinement as per the P6$_3$/mmc space group as can be seen from the excellent fit between the observed and calculated profiles shown in the inset (b) of **Figure 2** on a magnified scale. Details of Rietveld refinement and refined structural parameters are given in the supplementary information. The lattice constants obtained after Rieteveld refinement are a = b = 5.888(1) Å and c = 23.193(6) Å which are in excellent agreement with the values reported in the literature.[27] Consideration of the non-centrosymmetric space group P63mc proposed for Al$^{3+}$ doped BaFe$_{12}$O$_{19}$[28] did not lead to any improvement in the fits despite the fact that the number of refinemable parameters increased by 14. The chemical composition and hexagonal symmetry of the single-crystals were confirmed using energy-dispersive X-ray spectroscopy (EDX) and Laue diffraction measurements also and the details are given elsewhere.[29]

To capture departure from the Gorter model of the collinear magnetic structure of BFO, we used XMCD technique which probes the spin moment of the sample directly.[30–32] In BFO, partially filled d-orbitals of the Fe$^{3+}$ ions are responsible for spin moment. In the hexagonal unit cell of the P6$_3$/mmc space group of BFO, Fe$^{3+}$ ions are located in three different types of oxygen environments: trigonal bipyramid, tetrahedral and octahedral shown schematically in **Figure 1(b).** To check whether the spins are canted away from the c-axis, we obtained XMCD signals by measuring the XAS spectra using right circularly polarised (RCP) and left circularly polarised (LCP) beams in the grazing incidence (GI) geometry where the propagation vector **k** of the incident X-ray beam makes an angle of $15^0$ with respect to the c-axis of the crystal (see Fig. S1 of SI). The XAS spectra were recorded in total electron yield (TEY) mode without and with 0.5 T field applied in the direction of the X-ray beam. To get XMCD signals from the XAS spectra, we fit a linear background in the pre-edge region (700 eV – 704 eV) and then extrapolate it to full energy range (700 eV – 733 eV) and subtract it from the XAS spectra ($\sigma_+$) measured using the RCP X-ray beam. A similar linear background of the same slope was subtracted from XAS spectra ($\sigma_-$) measured using the LCP beam. XAS specra with linear background fit are shown in **Figure 3(a)**. After background subtraction, the post edge (730 eV – 733 eV) of $\sigma_+$ and $\sigma_-$ spectra were normalized to 1. Normalized XAS spectra thus obtained and their difference $\Delta\sigma = (\sigma_+ - \sigma_-)$, which represents the XMCD signal over the $L_{2,3}$ energy range, are depicted in **Figures 3(b)** and **3(c)**, respectively, for the GI geometry with no magnetic field (H = 0 T). The XMCD signal at the Fe $L_3$-edge shows a prominent at peak at ~ 709.5 eV which is known to be due to the $Fe^{3+}$ ion at the octahedral site.[33,34] The magnitude of the spin moment can be estimated using the well-established sum rule:[30]

$$m_{spin} + 7m_T^\theta = -\frac{(6P-4Q)}{R} n_h, \qquad \ldots\ldots.(1)$$

where $P = \int_{L_3}(\sigma_+ - \sigma_-)d\omega$ represents the integration of XMCD signal over the energy range of $L_3$-edge, $Q = \int_{L_3+L_2}(\sigma_+ - \sigma_-)d\omega$ represents the integration of the XMCD signal over the energy range of $L_{2,3}$-edges, and $R = \int_{L_3+L_2}(\sigma_+ + \sigma_-)d\omega$ represents the integration of the sum of XAS signal over the energy range of $L_{2,3}$-edges, $m_{spin}$ is the spin magnetic moment, $n_h$ the number of $Fe^{3+}$ 3d holes in BFO. $m_T^\theta$ is related to the magnetic dipole operator and its contribution is negligible for $3d^5 Fe^{3+}$.[34] To calculate spin magnetic moment using sum-rule, we followed the steps discussed in ref.[30] We first obtained the summed XAS spectra (= $\sigma_+$ + $\sigma_-$), then removed the edge jump using a two-step-like function and finally integrated the summed XAS spectra over the $L_{2,3}$ energy range. The variation of the XMCD signal integrated over the $L_{2,3}$ energy range and sum XAS with its integration are shown in **Figures 3c** and **3d,** respectively, for H = 0 T. We estimated the value of P, Q and R using these two figures and then calculated the spin moment using Equation (1). Normalized XAS spectra along with the XMCD signal measured in GI geometry without magnetic field (i.e., H = 0 T) and with a magnetic field of 0.5 T applied along **k**-vector, i.e., H//**k**, are compared in **Figures 4(a) and 4(b)**. It is evident from **Figure 4(a)** that the XMCD signal in the GI geometry in zero field shows a peak at ~ 709.5 eV while under a field of H = 0.5 T, a second peak at ~ 708.3 eV gets resolved **(see Figure 4(b)**. The peak at ~709.5 eV is attributed to $Fe^{3+}$ ion at the octahedral site while the small peak at ~708.3 eV is from $Fe^{3+}$ in trigonal bipyramid environment.[34] Using the P, Q and R values of GI geometry for H = 0 T and 0.5 T, the parallel and perpendicular components of the moment were calculated. The calculated spin moments parallel to the c-axis for the zero-field (H = 0 T) and H = 0.5 T field are found to be $m_{spin//c}$ ~ 0.2009 $\mu_B/Fe^{3+}$ and ~ 0.56228 $\mu_B/Fe^{3+}$, respectively, while those perpendicular to the c-axis are $m_{spin\perp c}$ ~ 0.0538 $\mu_B/Fe^{3+}$ and ~ 0.15058 $\mu_B/Fe^{3+}$, respectively. Application of magnetic field in the GI geometry obviously enhances the XMCD signal as well as $m_{spin\perp c}$ and $m_{spin//c}$ values as expected. The magnetic moment $m_{spin\perp c}$ clearly confirms the non-Gorter type non-collinear magnetic structure of BFO at 1.2 K raising doubts about the correctness of collinear magnetic structure of BFO assumed in a recent single-crystal neutron diffraction study at 4 K[22].

The canted nature of the $Fe^{3+}$ spins away from the c-axis in hexaferrites is known to lead to conical type magnetic order.[17,18,25,35] To verify this, we carried out single-crystal neutron diffraction studies at 1.5 K. As per the nuclear space group $P6_3/mmc$ of BFO, Bragg peaks for the *hhl* type reciprocal lattice rows with *l* = 2n ± 1 are forbidden. XRD pattern shown in **Figure 2** also confirms the absence of such reflections. These reflections are forbidden as per the collinear magnetic structure of the Gorter model with $P6_3/mm'c'$ magnetic space group also. The neutron diffraction pattern recorded along the *00l* reciprocal lattice row of BFO, however, shows some magnetic intensity around the *003* forbidden position (see **Figure 4(c)**) which clearly confirms the non-collinear nature of the spins in BFO in agreement with the XMCD results. A careful analysis of the forbidden peak observed around *003* position reveals that it is a triplet consisting of one peak at *003* position and two satellite peaks at *003* $\pm \tau$ (see the inset of **Figure 4(c)**). The origin of the two satellites at nearly symmetrical position about *003* reflection has been explained in terms of the block-type conical magnetic structure of hexaferrites[24,25] where the magnetic unit cell is divided into two blocks, R and R′, as depicted in **Figure 1(a)**. Within each block, the magnetic spins are canted away from the c-axis in such a way that the net magnetic moment is collinear within the block but noncollinear with respect to the neighbouring block. The block-type longitudinal conical magnetic structure // to c-axis is schematically illustrated in **Figure 1(c)** where the net moment of each block precesses about the c-axis and leads to the conical magnetic order with a half cone angle $\alpha$. The angle $\phi$ shown in **Figure 1(c)** is the phase difference between the ab-(basal) plane components of the net moments of the R and R′ blocks. It is given by $\phi = \pi(1-c/\tau)$.[23] The propagation vector ($k_0$) of the longitudinal conical structure is along the c-axis of the unit cell. The satellite peak in **Figure 4(b)** corresponds to $\tau = \pm 0.033$ which gives the periodicity (p ~ c/$\tau$) of the conical order as ~ 703 Å. This periodicity is nearly 30 times the value of the c-parameter of the nuclear unit cell. Since p is not exactly equal to 30c, the conical modulation of the BFO is incommensurate with respect to the collinear magnetic lattice. In case of Sc doped $BaFe_{12}O_{19}$ with x = 1.8, this periodicity is ~ 141 Å which is about 6 times the c-parameter.[23] The phase shift $\phi$ obtained by us for BFO is ~ $174^0$ as compared to $150^0$ for Sc doped BFO.[23] When the phase shift $\phi$ for the ab-components of the net moments of the R and R′ blocks becomes $180^0$, it leads to the appearance of screw-type spin structure[23] whose pitch is equal to the c-parameter of the nuclear unit cell. As a result of such an antiparallel arrangement of the ab-plane components, new magnetic reflections appear at $l = 2n \pm 1$ position along the *00l* reciprocal lattice row. However in the present case, the 003 peak is most likely due to multiple scattering as we do not see the 005 peak even though the two satellite peaks at *005 ± τ* are present. The presence of the two satellite peaks at *003 ± τ* shown in the inset of **Figure 4(c)** clearly confirms the conical type magnetic structure in the ground state of BFO.

To summarise, the evidence for non-collinear magnetic structure of BFO was presented using XAS and XMCD studies at 1.2 K. This was further confirmed by single-crystal neutron diffraction measurements at 1.5 K which reveal magnetic satellite peaks at *003 ± τ* positions which are forbidden by the nuclear $P6_3/mmc$ and magnetic $P6_3/mm'c'$ space groups for the collinear Gorter model. These satellite peaks are charecteristics of conical type magnetic structure. The periodicity of the conical order (~ 703 Å) is a non-integral multiple of the c-parameter suggesting incommensurate modulation. These observations provide the basis for the existence of magnetoelectric coupling under transverse field (H⊥c) in BFO also.

**Experimental Section**
Single-crystals of $BaFe_{12}O_{19}$ were grown by a high-temperature solution method, using sodium carbonate ($Na_2CO_3$) as flux for $BaFe_{12}O_{19}$ powder following the procedure described in ref.[36] The details of synthesis of $BaFe_{12}O_{19}$ powder are given in ref.[37] X-ray powder diffraction measurements were carried out using an 18-kW Cu-rotating anode-based powder diffractometer (Rigaku, model no. RINT 2500/PC series). Rietveld refinement was carried out using FullProf suite.[38] The XAS at Fe $L_{2,3}$-edges were recorded at the beam-line P04 of PETRA-III, Hamburg, Germany, using the low temperature setup described by Beeck et al.[39] Single-crystal neutron diffraction measurements were carried out using WISH time-of-flight diffractometer at ISIS, United Kingdom.[40] The resolution of the diffractometer ΔQ/Q is 0.3%.

**Supporting Information**
Supporting Information is available from the Wiley Online Library or from the author.

**Acknowledgments**
Portions of this research were carried out at the light source PETRA III of DESY, a member of the Helmholtz Association (HGF). Financial support from the Department of Science and Technology (DST), Government of India within the framework of the India@DESY is gratefully acknowledged. The neutron diffraction experiment on the single-crystal was performed using WISH diffractometer with the approval of the Science and Technology Facilities Council (STFC, proposal no. RB1968071) and was also funded by the DST.

Received: ((will be filled in by the editorial staff))
Revised: ((will be filled in by the editorial staff))
Published online: ((will be filled in by the editorial staff))

**Conflict of Interest**
The authors declare no conflict of interest.

# Evidence for Conical Magnetic Structure in M-type BaFe$_{12}$O$_{19}$ Hexaferrite: A Combined Single-Crystal XMCD and Neutron Diffraction Study

*Keshav Kumar, Shrawan K. Mishra, Sanjay Singh, Ivan Baev, Michael Martins, Fabio Orlandi, Pascal Manuel and Dhananjai Pandey\**

Mr. Keshav Kumar, Dr. Shrawan Kumar Mishra, Dr. Sanjay Singh, Prof. Dhananjai Pandey
School of Materials Science and Technology, Indian Institute of Technology (Banaras Hindu University), Varanasi, India-221005.
E-mail: dp.mst1979@gmail.com

Dr. Michael Martins, Dr. Ivan Baev
Universität Hamburg, Institut fü Experimentalphysik Luruper Chaussee 149, D-22761 Hamburg, Germany.

Dr. Pascal Manuel, Dr. Fabio Orlandi
Science and Technology Facilities Council Rutherford Appleton Laboratory ISIS Facility Didcot OX11 0QX, UK.

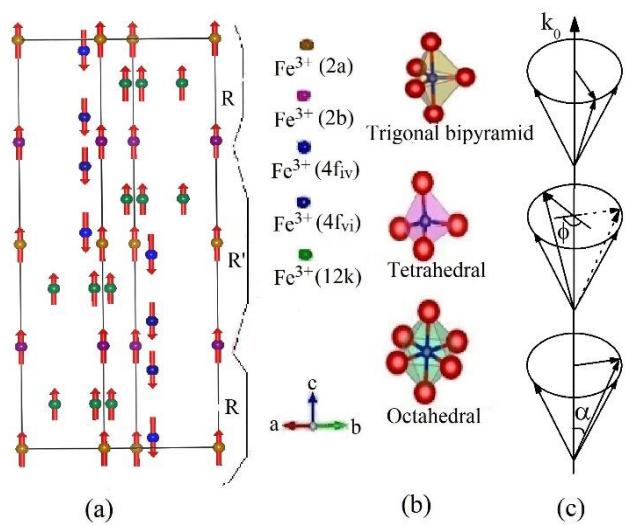

**Figure 01**: Schematic representation of (a) collinear magnetic structure of $BaFe_{12}O_{19}$, (b) coordination polyhedra for $Fe^{3+}$ ions in $BaFe_{12}O_{19}$ and (c) longitudinal conical magnetic structure due to precession of the net moment in the R and R′ blocks about the c-axis for canted spins.

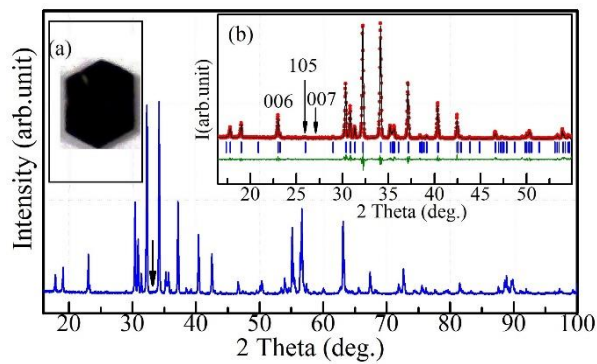

**Figure 02:** X-ray diffraction pattern collected on powder sample obtained after crushing $BaFe_{12}O_{19}$ crystals. The arrow shows the position of strongest Bragg reflection at $2\theta = 33.2^0$ for $\alpha$-$Fe_2O_3$. Inset (a) shows photograph of an as-grown crystal while (b) shows the results of Rietveld refinement where the observed, calculated, and difference profiles are shown with filled red circles, continuous black line and green line, respectively. The vertical blue bars in (b) represent the Bragg peak positions.

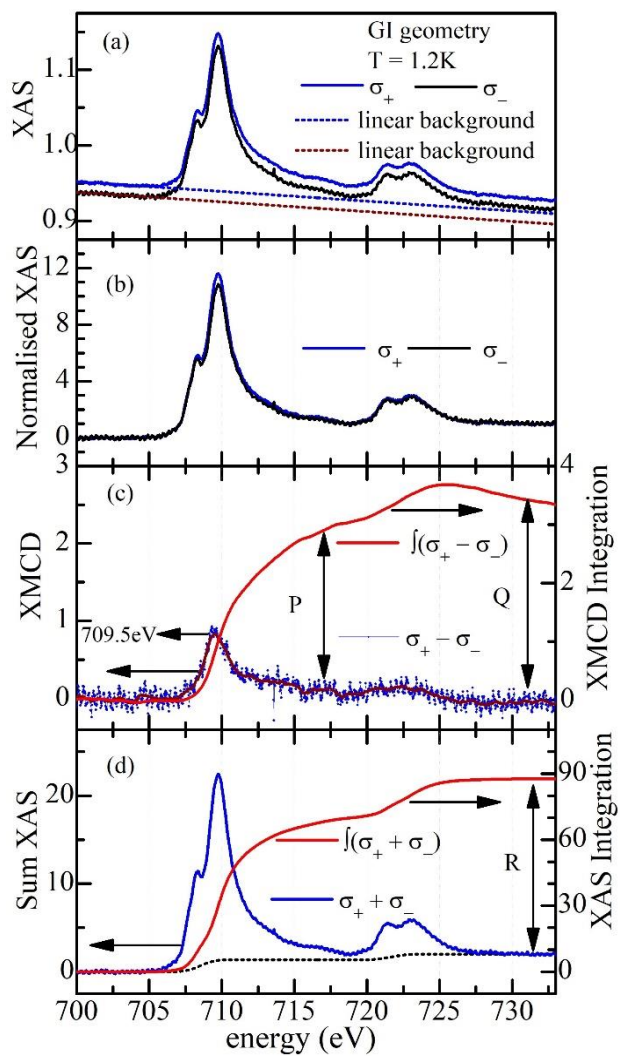

**Figure 03:** (a) XAS spectra recorded at Fe $L_{2,3}$-edges measured at 1.2 K on a single-crystal of $BaFe_{12}O_{19}$ in grazing incidence (GI) geometry using right ($\sigma_+$) and left ($\sigma_-$) circularly polarised X-ray beams. The dotted lines represent the linear background. (b) Normalised XAS spectra. The XMCD signal and summed XAS with their integration are shown in (c) and (d), respectively. The dotted line in (d) is the two-step-like function used for edge jump removal before XAS integration.

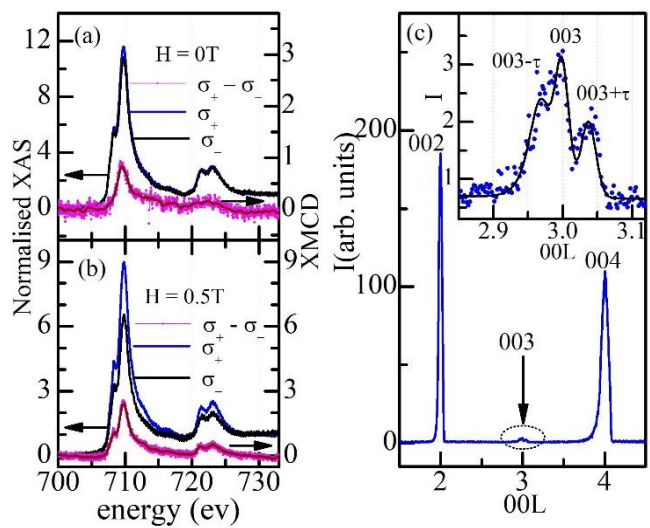

**Figure 04:** XAS spectra and XMCD signals of Fe at $L_{2,3}$-edges measured on a single-crystal of $BaFe_{12}O_{19}$ in grazing incidence geometry at 1.2 K (a) without and (b) with 0.5 T field. (c) Single-crystal neutron diffraction pattern recorded along *00l* reciprocal lattice row at 1.5 K depicting a magnetic peak around *003* position. The inset of (c) shows the triplet character of the magnetic peak with reflections at *l = 3* and *l = 3 ± τ* positions.